\documentclass[preprint2]{aastex}

\begin{document}

\title{Neutrino Spectra from Accretion Disks: Neutrino General Relativistic Effects and the Consequences for Nucleosynthesis}

\author{O. L. Caballero\altaffilmark{1}}
\affil{Department of Physics, North Carolina State University,
    Raleigh, NC 27695 \and Institute for Nuclear Theory, Seattle, WA 98195}
\author{G. C. McLaughlin\altaffilmark{2}}
\affil{Department of Physics, North Carolina State University,
    Raleigh, NC 27695}

\author{R. Surman\altaffilmark{3}}                        
\affil{Department of Physics and Astronomy, Union College, Schenectady, NY 12308}

\altaffiltext{1}{Electronic addresses: lcaballe@uw.edu, olcaball@ncsu.edu}
\altaffiltext{2}{Electronic address: gail\_mclaughlin@ncsu.edu}
\altaffiltext{3}{Electronic address: surmanr@union.edu}

\begin{abstract}
Black hole accretion disks have been proposed as good candidates for a range of interesting nucleosynthesis, including the $r$-process. The presence of the black 
hole influences the neutrino fluxes and affects the nucleosynthesis resulting from the interaction of the emitted neutrinos and hot outflowing material ejected 
from the disk.  We study the impact of general relativistic effects on the neutrinos emitted from black hole accretion disks.  We present abundances obtained by 
considering null geodesics and energy shifts for two different disk models. We find that both the bending of the neutrino trajectories and the energy shifts have 
important consequences for the nucleosynthetic outcome.
\end{abstract}


\section{Introduction}

Rapidly accreting disks around black holes arise in several scenarios.  Disks with MeV or higher temperatures have been discussed in the context of compact object 
mergers, long and short duration gamma ray bursts, and core collapse supernovae which have a rotating progenitor.

The coalescence of compact objects produces hot disks \citep{Lee1999, Rosswog2004, Taniguchi2005}, and the evolution of these objects is of great importance for 
several reasons.  Mergers generate gravitational waves susceptible to detection with interferometric gravitational wave detectors, e.g. \citet{Abbott} and 
references therein.  Energy release from these objects has been suggested as a promising source for the generation of gamma ray bursts \citep{Popham1999, 
RuffertGRB-BH, SetiawanBHmerger, Kneller:2004}.  The neutrino flux from these objects is so large, that it would be easily detected by currently online neutrino detectors \citep{McLaughlin07,Caballerosurface}.

These objects will also eject nucleosynthetic products that must be considered when studying the galactic inventory of elements. Thus it is important to determine 
the type of elements that are produced from accretion disks.  Collisions have been speculated to contribute to the synthesis of neutron rich nuclei via the 
$r$-process, due to the decompression of cold or mildly heated neutron star matter during the merger \citep{Lattimer1976,Meyer, Goriely,Freiburghaus, Metzger10}.  
It has also been suggeted that they form $r$-process nuclei from the ejection of hot material during the merger \citep{McLaughlin, Oechslin, Surman08, Metzger08, 
Wanajo, Kajino10}.

\citet{McLaughlin} and \citet{Surman08} found neutron rich material in the outflow ejected out of the plane of the inner hot regions of an accretion disk. 
A successful $r$-process occurred when either (1) low entropy and fast outflow conditions obtained or (2) a favorable combination of electron neutrino and anti-neutrino 
spectra, which when taken together with the accretion disk geometry allowed the material to flow far away from the source of the neutrinos before nuclei begin 
to form.  Disks can form other elements as well.  Disks that are expected to occur in rare core collapse supernovae. e.g. ``collapsars'' \citep{MacFadyen1999} are 
dominated by trapped electron neutrinos or are mostly transparent.  This leads to elements that are formed when roughly equal numbers of neutrons and protons are 
present, such as Nickel-56 and $p$-process elements \citep{Surman06,Pruet,Kizivat}.

The presence of a massive object such as a black hole changes the properties of space-time around it affecting the spectra of radiation emerging from the matter 
located in its vicinity. Since the neutrino spectra are crucial to the outcome of the nucleosynthesis in hot outflows \citep{Meyer95, McLaughlin96, McLaughlin96a, 
Haxton, Meyer98, Frohlich}, particularly in accretion disk hot outflows \citep{Surman03}, it is important to consider the effects of general relativity on their 
energies and trajectories.

Several detailed studies have been conducted for photons that emerge from accretion disks and these results are also applicable to neutrinos.  In an early work 
\citet{Luminet} studied the photon spectrum from black hole accretion disks as seen by an observer located at infinity. The effects of light bending and energy 
shifts for photons emerging from a Schwarzschild black hole were included. \citet{Luminet} also considered flat disks and observers located at different 
inclination angles with respect to the axis of symmetry of the disk. \citet{Fukue} studied the spectra of a non-monochromatic distribution by placing the observer 
at an arbitrary distance from the black hole.  \citet{Cunningham} calculated the X-ray spectrum emerging from a disk around a Kerr black hole.  
\citet{Bhattacharyya} studied the relativistic spectra when the black hole is replaced by a rapidly rotating neutron star. A more recent work compared different 
models for the accretion disk and applied their results to ultraluminous X-ray sources \citep{Lorenzin}. To date, studies of the effects of general relativity, 
including ray bending, on neutrinos emerging from accretion disks have focused on neutrino pair annihilation rates and their production of gamma ray bursts 
\citep{Asano, Birk, Zalamea}.

We consider neutrino general relativistic effects with a focus on the consequences for the production of nuclei in the vicinity of accretion disks. We are 
interested on the nucleosynthesis occurring when hot winds of free nucleons ejected from the disk interact with the neutrinos emitted by the disk. In this paper we 
incorporate effects such as neutrino bending and energy shifts to the neutrino spectra. Our study includes two accretion disk models: a dynamical model coming from 
3D simulations and a steady state disk in one dimension. 

This paper is organized as follows: in Sec. \ref{general relativistic effects} we present the basic formalism used to determine neutrino ray bending, energy shifts 
and neutrino fluxes. In Sec. \ref{nucleosynthesis} we introduce the disk models and describe our outflow model. In Sec. \ref{results} we discuss the 
nucleosynthetic outcomes for various scenarios and in Sec. \ref{conclusions} we conclude.

%==========================================================================
\section{General Relativistic effects}
\label{general relativistic effects}
Our treatment of general relativistic effects is divided in two main components: bending of neutrino trajectories and energy shifts. For the last effect we have 
found energy shifts in the Kerr and Schwarzschild metrics. However, for simplicity, we use only the Schwarzschild metric for our treatment of the neutrino 
trajectories. In the next two subsections we describe our approaches to each of these effects. We label the points from where neutrinos are emitted as $r_{em}$ and 
the points where the neutrinos fluxes are observed as $r_{ob}$. The emission points, $r_{em}$, correspond points on the neutrino surfaces (found in a previous work 
\citep{Caballerosurface}) and $r_{ob}$ to all the points on the outflow trajectory where nuclear products are synthesized.

\subsection{Neutrino trajectories}\label{neutrinotrajectories}
Neutrino trajectories correspond to null geodesics in the curved space. Here we use the same notation and follow a similar methodology as in \citet{Muller} to find 
neutrino trajectories. Our main objective is to find the angles $\xi$ and $\alpha$ that a neutrino forms on the sky of an observer located at a distance $r_{ob}$ 
from a black hole.  The angle $\xi$ describes the direction of momentum of the neutrino with respect to line joining the observer and the black hole, while 
$\alpha$ is the angle with respect to a horizontal, both directions defined in a coordinate system centered at the observer. The angles $\xi$ and $\alpha$ are found after 
solving the equation of motion for a neutrino in the vicinity of a black hole.

We start by describing the neutrino trajectory in a coordinate frame centered at the black hole. Then we find the required variables in the observer's coordinate 
system. For simplicity we assume that the space time around the black hole is described by the Schwarzschild metric,
\begin{eqnarray}
ds^2&=&-\left(1-\frac{r_s}{r}\right)dt^2+\frac{1}{1-r_s/r}dr^2 \nonumber \\
& & + r^2\left(d\theta^2+\sin^2\theta d\varphi^2\right),
\label{smetric}
\end{eqnarray} 
where $r_s=2M$ and $M$ is the black hole mass. The spherical symmetry of the Schwarzschild metric makes it possible to solve the equations of motion in the 
$\theta=\pi/2$ plane. In this way the Lagrangian formalism results in null geodesics described by
\begin{equation}
\left[\frac{1}{r^2}\left(\frac{dr}{d\varphi}\right)\right]^2+\frac{1}{r^2}\left(1-r_s/r\right)=\frac{1}{b^2},
\end{equation}
or
\begin{equation}
\int^{\varphi_{ob}}_{\varphi_{em}}d\varphi=\pm\displaystyle\int^{r_{ob}}_{r_{em}}\frac{dr}{r\sqrt{\frac{r^2}{b^2}-\left(1-\frac{r_s}{r}\right)}}.
\label{trajectoryeq}
\end{equation}
The sign of equation~\ref{trajectoryeq} depends on whether the neutrino is approaching or leaving the black hole. In the above equation $b$, the impact parameter 
at infinity, is the ratio between the neutrino energy $E$ and the its angular momentum $L$, $b=L/E$. The impact parameter $b$ is a constant along the neutrino 
trajectory and makes it possible to calculate the angle $\xi$ between the neutrino trajectory and the radial direction at any point $r$ by
\begin{equation}
b=\frac{r \sin\xi}{\sqrt{1-r_s/r}}.
\label{xi}
\end{equation}
A neutrino moves radially if $b=0$ and tangentially if $\xi=\pi/2$. At the point of closest approach to the black hole $r_+$,
\begin{equation}
b=\frac{r_+ }{\sqrt{1-r_s/r_+}}.
\label{closest}
\end{equation}

At $r_+$, the largest root of the square root in equation \ref{trajectoryeq}, there is inversion of the radial motion. It corresponds to instantaneous tangential 
motion. Also, there exits a critical impact parameter $b_{crit}=\sqrt{3}r_{crit}$, with $r_{crit}=3r_s/2$ such that null geodesics with $b<b_{crit}$ are captured 
by the black hole.

\begin{figure}[h]
%\epsscale{1}
\plotone{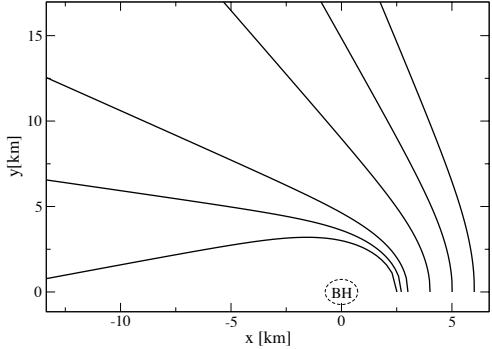}
\caption{Null geodesics, found by using equation \ref{trajectoryeq}, starting at $\varphi_{em}=0^\circ$ and distances $x=r_{em}= r_+=2.5, 2.7, 3, 4, 5, 6$ km. The black hole is located at ($x=0, y=0$) and its mass is $0.5M_\odot$. The impact parameter $b$ is found using equation \ref{closest}.\label{geodesics}}
\end{figure}

We can use equation \ref{trajectoryeq} to our convenience. For example, we can start from an initial emission point ($r_{em}, 
\varphi_{em}$) and a known impact parameter $b$ and draw neutrino trajectories starting from that point. Figure \ref{geodesics} 
shows null geodesics followed by neutrinos leaving at $\varphi_{em}=0^\circ$ and different distances $r_{em}=r_+$ from a black 
hole. The impact parameter $b$ is calculated using equation \ref{closest}.

\begin{figure}[h]
\epsscale{1.05}
\plotone{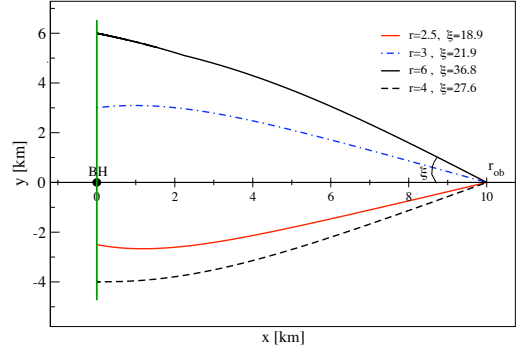}
\caption{Null geodesics, found by using equation \ref{trajectoryeq}, starting from the vertical line at different distances $r$ from a black hole (centered dot) and ending at the same observation point $r_{ob}=10$ km, $\varphi_{ob}=0$. The black hole mass is $0.5M_\odot$. $\xi$ is the angle formed line joining the black hole and the observer and the direction of the null geodesic at $r_{ob}$.\label{rays}}
\end{figure}

\begin{figure}[h]
\epsscale{1.05}
\plotone{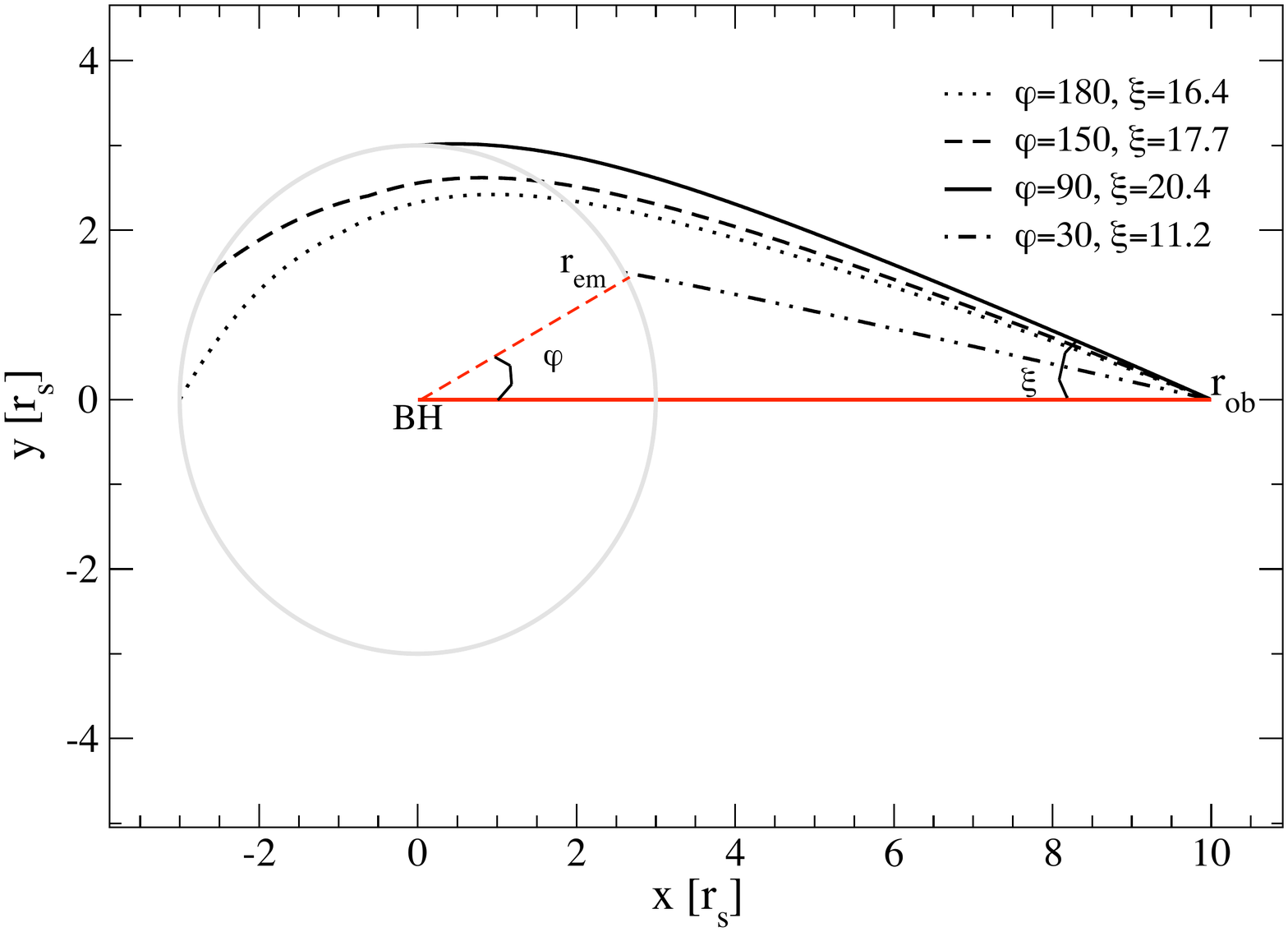}
\caption{Null geodesics starting at different angles $\varphi$, but from the same distances $r_{em}=3r_s$ to the black hole (BH) located at ($x=y=0$). The black hole mass is $0.5M\odot$ and $r_s=2.9$ km. The observer is at $r_{ob}$=10$r_s$. \label{rays2}}
\end{figure}

We can also fix the initial and final points of the trajectory, which are the limits of the integrals in equation \ref{trajectoryeq}, and find the impact parameter 
$b$ that satisfies it. By solving for $b$ we are also finding the angle $\xi$ that the null geodesic makes at any point of the trajectory by using equation 
\ref{xi}. Depending on the position of the emitter and the observer with respect to the black hole this could imply solving consistently equation \ref{closest} to 
obtain the point of closest approach. Figures \ref{rays} and \ref{rays2} show different trajectories for which we have found $\xi$ using the procedure described 
above.

In the present study the distance from the emission points on the neutrino surface to the black hole are such that $r_\nu=r_{em}>r_{crit}$. Therefore we are not 
concerned about geodesics that end up into the black hole. We are also interested in null geodesics that start from a fixed point on the neutrino surface and end 
at another fixed point on the outflow.  In such case we focus on finding $b$ from equation \ref{trajectoryeq} depending on the initial conditions $r_{em}, 
\varphi_{em}$ and $r_{ob}, \varphi_{ob}$. Taking into account the relationship between the distances $r_{em}$ and $r_{ob}$, our study requires the solution of the 
following cases, which are sketched in Figure \ref{cases} (a study of more cases in the strong deflection limit can be found in \citet{Bozza}):

\begin{figure}[h]
\epsscale{1.}
\plotone{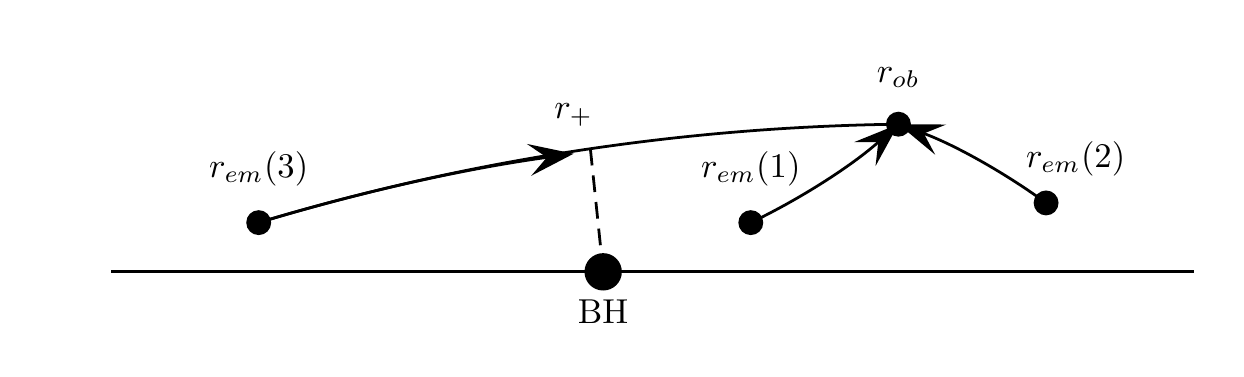}
\caption{Different relationships between the distances from the emission point $r_{em}$  to the black hole (BH) and from the observation point $r_{ob}$ to the BH. $r_+$ represents the point of closest approach to the BH.The labels (1), (2), and (3) represent the cases discussed in the text. \label{cases}}
\end{figure}

{\it Case 1}. The distance from the observer to the black hole is larger than the distance from the emission point to the black hole, $r_{ob}>r_{em}$, and the 
neutrino does not reach the point of closest approach, $r_{em}>r_+$. In this case neutrinos are leaving the black hole and we need to solve equation 
\ref{trajectoryeq} with the positive sign.

{\it Case 2}. The distance from the emission point to the black hole is larger than the distance from the observation point to the black hole, $r_{em}>r_{ob}$, and 
the neutrino does not reach the point of closest approach to the black hole, $r_{ob}>r_+$. In this case the neutrinos are approaching the black hole, so we need to 
solve equation \ref{trajectoryeq} with the negative sign.
\begin{equation}
\int^{\varphi_{ob}}_{\varphi_{em}}d\varphi=-\int^{r_{ob}}_{r_{em}}\frac{dr}{r\sqrt{\frac{r^2}{b^2}-(1-\frac{r_s}{r})}}.
\end{equation}

{\it Case 3}. The distance from the emission point to the black hole is larger than the distance from the observer to the black hole and the neutrino trajectory 
reaches the point of closest approach to the black hole. In this case we need to split equation \ref{trajectoryeq} in two branches. One is the approaching phase 
(neutrinos leave the neutrino surface and approach to $r_+$), and the departure phase in which neutrinos leave $r_+$ to get to the observer. We assign the signs in 
equation \ref{trajectoryeq} accordingly to get
\begin{eqnarray}
\int^{\varphi_{ob}}_{\varphi_{em}}d\varphi&=&-\int^{r_+}_{r_{em}}\frac{dr}{r\sqrt{\frac{r^2}{b^2}-(1-\frac{r_s}{r})}} \nonumber \\
 & & + \int^{r_{ob}}_{r_+}\frac{dr}{r\sqrt{\frac{r^2}{b^2}-(1-\frac{r_s}{r})}}.
\end{eqnarray}

This last case is similar to the situation where $r_{em}<r_{ob}$ and $r_+$ is reached. For this latter situation, one would switch $r_{ob}$ and $r_{em}$ for case 3 
in Figure \ref{cases}.

Whether the neutrino reaches $r_+$ or not is not known a priori. Therefore we need to consider all the cases when finding null geodesics for our problem. We use a 
root finding algorithm to solve the corresponding cases of equation \ref{trajectoryeq}. We find values of $b$ and $r_+$ such that equation \ref{closest} is valid 
and consistent with the appropriate case.

\begin{figure}[h]
\epsscale{1.25}
\plotone{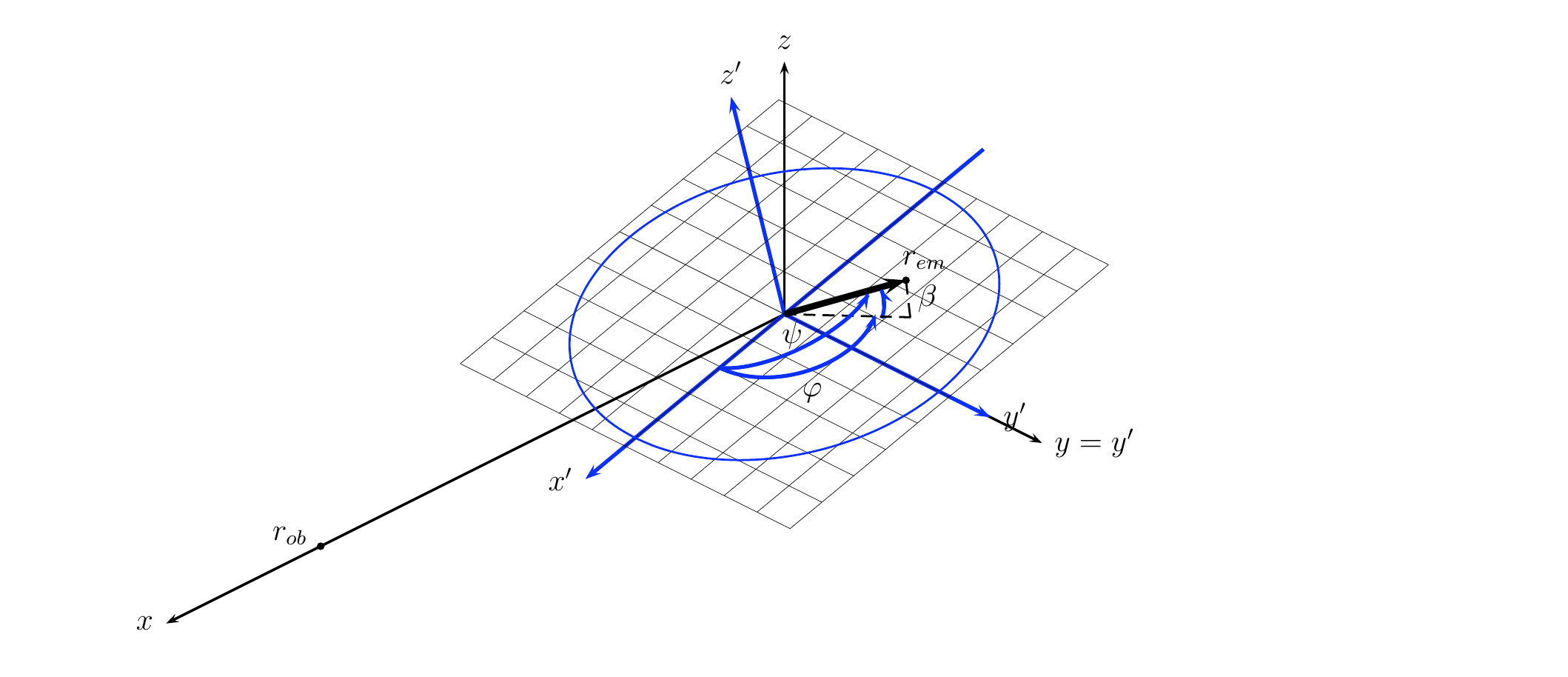}
\caption{A neutrino is emitted from the point $r_{em}$, which is at some distance above the plane of the disk (circle). The black hole is at the center of the disk. The $x',y',z'$ correspond to the disk system. $z'$ is normal to the disk. With respect to this system the emission point has coordinates $(r_{em},\varphi,\theta=\pi/2-\beta)$. $\psi$ is the angle between the $x'$-axis and a line joining the emission point to the black hole. \label{axes}}
\end{figure}

\begin{figure}[h]
\epsscale{1.25}
\plotone{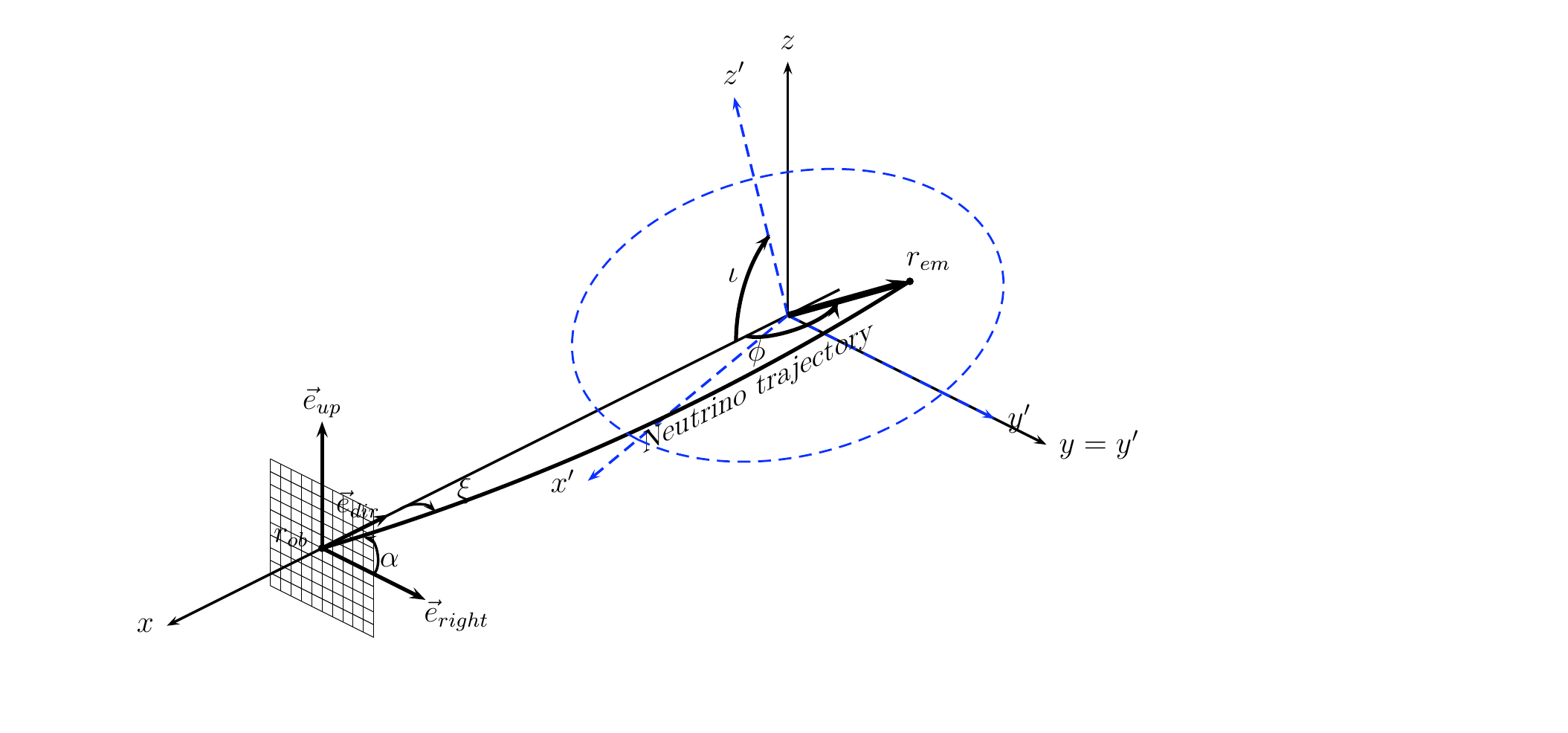}
\caption{Same coordinate systems as in Figure \ref{axes}. The angle between the axis normal to the disk, $z'$, and the observer direction $x$, is $\iota$. Here the coordinates of the emission point are given as seen from the $x,y,z$ system. The neutrino trajectory forms a plane with the $x$-axis. The neutrino hits the observer plane and forms and angle $\alpha$ with respect to the $\vec{e}_{right}$ direction and an angle $\xi$ with respect to $\vec{e}_{dir}$. \label{axes2}}
\end{figure}

Now we proceed to determine the angles that a neutrino describes on the observer's sky. Our purpose is to find $\alpha$ and $\xi$ as seen at $r_{ob}$. We place the 
observer at an arbitrary angle above the plane of Figure \ref{rays}. Figure \ref{axes} shows the geometry and variables we will use to discuss the problem. The 
disk is represented by the circle.  We define three coordinate systems. One corresponds to the disk ($x',y',z'$), the $z'$-axis is normal to plane of the disk. The second 
reference ($x,y,z$) is centered at the disk but aligned in such a way that the observer is located on the $x$-axis at a distance $r_{ob}$ and $y=y'$. The point of 
observation $r_{ob}$ has an inclination $\iota\in(0,\pi/2) $ with respect to the normal to the disk (see Figure \ref{axes2}).

The relation between these the coordinate frames is given by a rotation around the $y$-axis:
\begin{eqnarray}
\vec{e}_{x'}&=&(\sin\iota, 0, -\cos\iota),\\
\vec{e}_{y'}&=&\vec{e}_y,\\
\vec{e}_{z'}&=&(\cos\iota, 0,\sin\iota).
\end{eqnarray}

In Figure \ref{axes2} we show the third coordinate system which corresponds to the observer reference frame and is just a translation of the $x,y,z$ reference with 
origin at $r_{ob}$. In this case $\vec{e}_{dir}=-\hat{x}$, $\vec{e}_{right}=\hat{y}$, $\vec{e}_{up}=\hat{z}$. Note that $\vec{e}_{dir}$ points towards the black 
hole.

An emitted neutrino has coordinates $(r_{em},\varphi,\theta=\pi/2-\beta)$ with respect to the $x',y',z'$ axes (Figure \ref{axes}). The angle between the $x'$-axis 
and $r_{em}$ is $\psi$. On the other hand, on the $x,y,z$ axes, the $x$-axis and the emission point form an angle $\phi_{em}$ (Figure \ref{axes2}). The neutrino 
trajectories are restricted to a plane, called the observational plane, because of the spherical symmetry of space time in the Schwarzschild metric. This plane is 
defined by the null geodesic traveling from the emission point to the observer and the $x$-axis. We therefore can solve equation \ref{trajectoryeq} on that plane, 
which corresponds to solving for $\alpha$ and $\xi$ with initial conditions $r_{em}$ and $\phi_{em}$.
 
\begin{figure}[h!]
\epsscale{.7}
\plotone{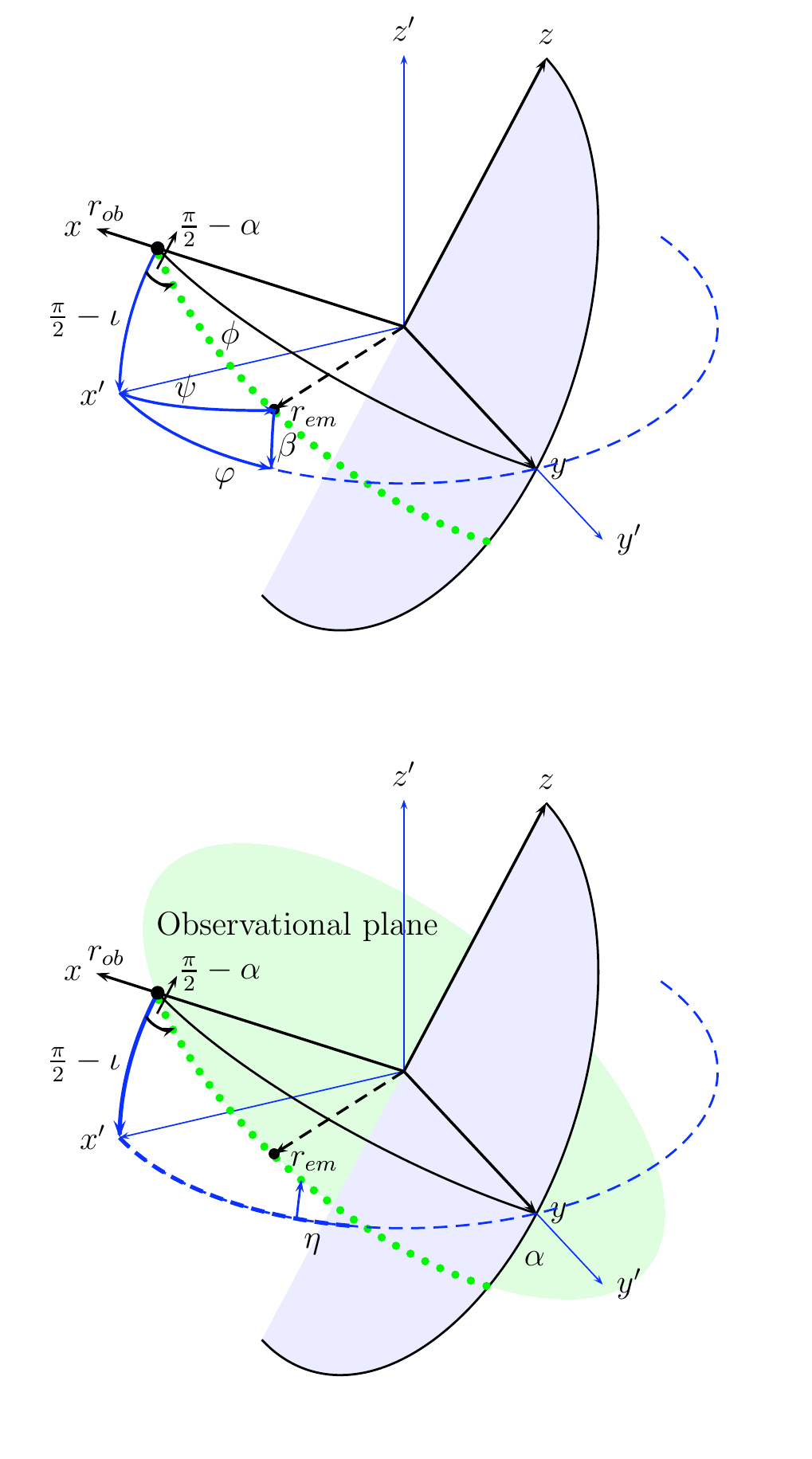}
\caption{Same coordinate systems as in Figure \ref{axes}. Here we have changed the point of view in such a way that the observer's system seems rotated with respect to the disk system. The blue dashed circle represents the equatorial plane of the disk. The angles $\varphi$, $\phi$, $\psi$, $\alpha$ and $\iota$ are as defined in Figures \ref{axes} and \ref{axes2}. A neutrino is emitted at the $r_{em}$ and arrives to the point $r_{ob}$. The neutrino trajectory (green dotted line) belongs to the observational plane. The observational plane forms an angle $\alpha$ with the $y$ axis in the $yz$ plane. Top: a relationship for $\phi$ is found by using the spherical triangles described by the arcs $\psi$, $\phi$ and $\frac{\pi}{2}-\iota$; and $\beta$, $\varphi$ and $\psi$. Bottom: the neutrino trajectory forms and angle $\eta$ with the plane $x'y'$. Using the spherical triangle described by the observational plane, the $x'y'$ and the $x'z'$  planes we find a relationship for $\eta$ in terms of $\alpha$ and $\iota$. \label{angles}}
\end{figure}

An image plane (as seen from the observer's reference, i.e. consisting of the $xy$ plane in Figures \ref{axes} and \ref{axes2}) will intersect the observational plane 
at an angle $\alpha$ with respect to $\vec{e}_{right}$, or equivalently the $y$-axis. The angle $\alpha$ can be obtained from the relationships between spherical 
triangles of Figure \ref{angles} (top). In this figure, the neutrino is emitted at $r_{em}$ and follows a trajectory represented by the dotted line reaching the 
observer at $r_{ob}$. The angle between the neutrino observational plane and the $xz$ plane is $\frac{\pi}{2}-\alpha$. Using the spherical triangle defined by the 
arcs $\psi$, $\phi$ and $\frac{\pi}{2}-\iota$, and the spherical triangle defined by the arcs $\psi$, $\phi$ and $\beta$ we find
\begin{equation}
\cos\alpha=\frac{\sin\varphi\cos\beta}{\sin\phi}.
\end{equation}
Transforming the coordinate $x'$ of the emission point to the $x$ axis, after a rotation around the $y$ axis by the angle $\iota$, we find
\begin{eqnarray}
\cos\phi&=&\sin\iota\cos\psi+\cos\iota\sin\beta\\
\sin\phi&=&\sqrt{1-\cos^2\phi}%=\sin\psi\sqrt{1+\cot^2\psi\cos^2\iota},
\end{eqnarray}
where $\cos\psi=\cos\beta\cos\varphi$.

Solving for the different cases of equation \ref{trajectoryeq} we find the unique geodesic that connects the point ($r_{em},\varphi_{em}=\varphi$) to 
($r_{ob},\phi_{ob}=0$). This means we find $b$ and $\alpha$ and by virtue of equation \ref{xi}, we find $\xi$. We therefore have a relation between 
$(r_{em},\varphi)$ and $(\xi, \alpha)$. Once $\xi$ and $\alpha$ are found for every point over the neutrino surface, we can calculate the solid angle 
$d\Omega_{ob}=\sin\xi d\xi d\alpha$ covered by the observer.

It is also useful to find a relationship for the angle $\eta$ that the neutrino trajectory makes with the plane of the disk in terms of the already known angles. 
We find $\eta$ by using the law of cosines for the spherical triangle described by the neutrino trajectory, the disk and the $xz$ planes (see Figure \ref{angles} 
(bottom))
\begin{equation}
\cos\eta=\cos\alpha\sin\iota.
\end{equation}
%======================================================================
\subsection{Energy red-shift}
\label{energy red-shift}
The energy measured by an observer $E_{ob}$ changes from the emitted energy  $E_{em}$ by the red-shift factor $(1+z)$,
\begin{equation}
\frac{E_{em}}{E_{ob}}=1+z.
\end{equation}

In the case of a massless neutrino its measured energy is the projection of the neutrino 4-momentum $p$ on the 4-velocity $u$ of the emitting matter, $p_\beta 
u^\beta$. Then a general expression for the red-shift is,
\begin{equation}
1+z=\frac{\left(p_tu^t+p_ru^r+p_\theta u^\theta+p_\varphi u^\varphi\right)_{em}}{\left(p_tu^t+p_ru^r+p_\theta u^\theta+p_\varphi u^\varphi\right)_{ob}}.
\label{generalshift}
\end{equation}
 In equation \ref{generalshift} (and in the next discussions) the subindexes $em$ and $ob$ indicate that the components of the metric tensor 
(or any other quantity) should be computed with the observer or emitter coordinates accordingly.  From this general expression we can consider specific cases according to the movement of the emitting particle and the metric of the space-time. For example, if 
the emitter is considered at rest then $u^\varphi=u^\theta=u^r=0$. For a cloud of nonrotating gas accreted by a black hole we have $u^\varphi=u^\theta$=0, while 
$u^r\ne0$.  We could also have a case where the emitting and the receiving particles do not have movement in the $r$ and $\theta$ directions, and therefore 
$u^r=u^\theta=0$ but $u^\varphi\ne0$. In what follows we study different cases depending on the emitter motion and the space-time metric.

\subsubsection{Non-rotating black hole}

For a non-rotating black hole we consider here two cases: in the first one the emitter and the observer do not have relative motion, and in the second case both 
rotate around the black hole. In the first case $u^\varphi=u^\theta=u^r=0$. Then the general expression of equation \ref{generalshift} becomes
\begin{equation}
1+z=\frac{p^{em}_tu_{em}^t} {p^{ob}_t u_{ob}^t},
\end{equation}
where $p_t$ is the neutrino energy which is constant along the neutrino trajectories. Therefore we have
\begin{equation}
1+z=\frac{u_{em}^t} {u_{ob}^t},
\end{equation}
where $u^t=dt/d\tau$. $d\tau$ is calculated according to the metric describing the curvature of the space-time,
\begin{equation}
d\tau=\left(-g_{\alpha\beta} dx^\alpha dx^\beta\right)^{1/2}.
\end{equation}

If the space-time around the black hole can be described by the Schwarzschild metric (equation \ref{smetric}) then there is spherical symmetry, and therefore 
$d\tau=(-g_{tt})^{1/2}dt$.  We then find the well known expression for the energy red-shift,
\begin{equation}
1+z=\frac{(-g_{tt})^{1/2}_{ob}}{(-g_{tt})^{1/2}_{em}},
\label{sshift}
\end{equation}
where $g_{tt}=-(1-r_s/r)$.

\begin{figure}[h]
\epsscale{0.7}
\plotone{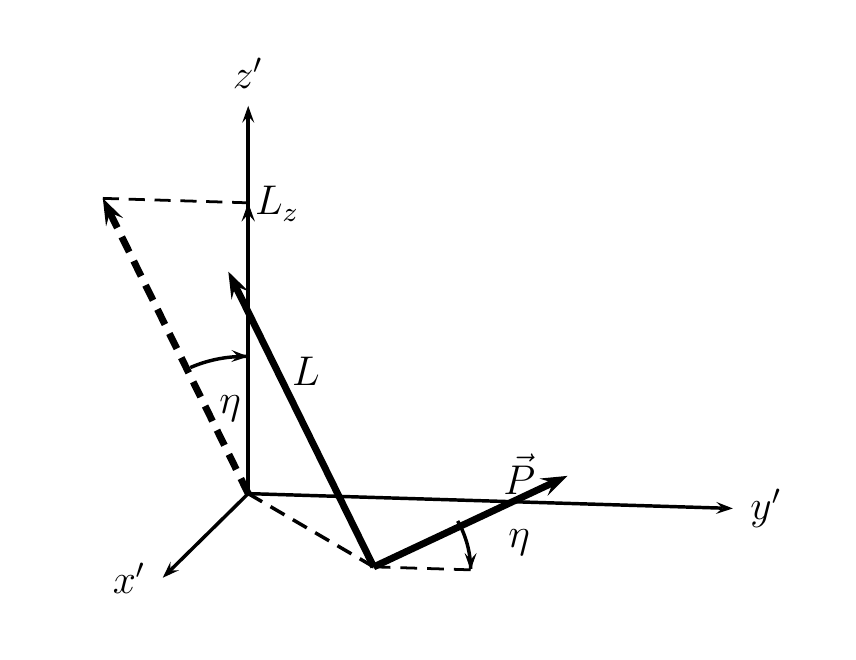}
\caption{A neutrino is emitted with momentum $\vec{P}$. The neutrino trajectory and momentum form an angle $\eta$ with the $x'y'$ plane. The $z^\prime$ component of the angular momentum is $L_{z^\prime}=L\cos\eta$. \label{momentum}}
\end{figure}

In the second case we are considering here, the emitter and the observer are rotating in stationary orbits around the black hole with angular velocities 
$\Omega_{em}$ and $\Omega_{ob}$, respectively. This case exemplifies a particle that has a stationary orbit around the black hole and is located at a fixed angle 
$\theta_{em}$. This particular choice could also represent elements of matter of a steady state accretion disk. In Keplerian rotation the values for the angular 
velocity for the observer or emitter $\Omega$ is
\begin{equation}
\Omega=\frac{M^{1/2}}{r^{3/2}}.
\end{equation}
In this case we have for the radial and azimuthal velocities $u^r=u^\theta=0$, and therefore the general expression for the red-shift (equation \ref{generalshift}) 
is reduced to
\begin{equation}
1+z=\frac{p^{em}_tu_{em}^t\left(1+\frac{p^{em}_\varphi} {p^{em}_t} \frac{u_{em}^\varphi}{u_{em}^t}\right)}{p^{ob}_tu_{ob}^t\left(1+\frac{p^{ob}_\varphi} {p^{ob}_t} \frac{u_{ob}^\varphi}{u_{ob}^t}\right)}.
\end{equation}
The second fraction in the parenthesis equals the angular velocity $\Omega=u^\varphi/u^t=d\varphi/dt$. Furthermore, $p_\varphi$ and $p_t$ are constants along the 
neutrino trajectories, which correspond to the projection of the angular momentum on the $z'$-axis $L_{z'}$ and energy $E$ respectively. In the previous section 
we introduced the impact parameter $b=L/E$. Using these quantities we re-write the red-shift as
\begin{equation}
1+z=\frac{u^t_{em}\left(1+\Omega_{em}b\cos\eta\right)} {u_{ob}^t\left(1+ \Omega_{ob}b\cos\eta\right)},
\end{equation}
where $\eta$ is the angle formed by $p_\varphi$ and the plane of the disk $\theta=\pi/2$ (see Figure \ref{momentum}).

Using again the fact that $u^t=dt/d\tau$, we get an expression for the energy shift in the Schwarzschild metric, when both observer and emitter rotate around the 
black hole (this derivation can be found in \citep{Luminet} or \citep{Thorne}),
\begin{equation}
1+z=\frac{(-g_{tt})^{1/2}_{ob}\left(1+\Omega_{em}b\cos\eta\right)} {(-g_{tt})^{1/2}_{em}\left(1+ \Omega_{ob}b\cos\eta\right)}.
\label{sshiftrot}
\end{equation}

\subsubsection{Rotating black hole}
In the case of a non-charged, rotating black hole the curvature of the line element  can be written in the Kerr geometry as, \citep{gravitation},
\begin{equation}
ds^2=g_{tt}dt^2+2g_{t\varphi}dtd\varphi +g_{\varphi\varphi}d\varphi^2 +g_{rr}dr^2+g_{\theta\theta}d\theta^2,
\end{equation}
with
\begin{mathletters}
\begin{eqnarray}
g_{tt}&=&-\left(1-\frac{r_sr}{\Upsilon^2}\right),\label{kmetrictensor}\\
g_{rr}&=&\Upsilon^2/\Delta,\\
g_{t\varphi}&=&-\frac{r_sra\sin^2\theta}{\Upsilon^2},\\
g_{\varphi\varphi}&=&\frac{\sin^2\theta}{\Upsilon^2}\left[(r^2+a^2)^2-\Delta a^2\sin^2\theta\right],\\
g_{\theta\theta}&=&\Upsilon^2.
\end{eqnarray}
\end{mathletters}
Here $a=Jc/GM^2$ is the spin of the black hole ($J$ is the total angular momentum), and $\Delta$ and $\Upsilon$ are given by
\begin{mathletters}
\begin{eqnarray}
\Delta=r^2-r_sr+a^2,\\
\Upsilon^2=r^2+a^2\cos^2\theta.
\end{eqnarray}
\end{mathletters}

Considering the first case in which the emitter and observer have no relative motion we have a similar expression for the the red-shift as in equation \ref{sshift} 
but in this case with $g_{tt}$ defined as in equation \ref{kmetrictensor}.
\begin{equation}
1+z=\frac{(-g_{tt})^{1/2}_{ob}} {(-g_{tt})^{1/2}_{em}}=\frac{\left(1-\frac{r_sr}{\Upsilon^2}\right)^{1/2}_{ob}}{\left(1-\frac{r_sr}{\Upsilon^2}\right)^{1/2}_{em}}.
\label{kshift}
\end{equation}

Now, as in the previous subsection, we consider the case where the observer and the emitter rotate around the black hole. The conditions for the 4-velocities 
($u^r=u^\theta=0$) and the metric of space-time lead to
\begin{equation}
d\tau=\left[dt^2\left(-g_{tt}-2g_{t\varphi}\frac{d\varphi}{dt}-g_{\varphi\varphi}\frac{d\varphi^2}{dt^2}\right)\right]^{1/2},
\end{equation}
or in terms of the angular velocity, 
\begin{equation}
\frac{dt}{d\tau}=\frac{1}{\left(-g_{tt}-2g_{t\varphi}\Omega-g_{\varphi\varphi}\Omega^2\right)^{1/2}}.
\end{equation}
Therefore the red-shift becomes,
\begin{eqnarray}
1+z&=&\frac{\left[-g_{tt}-2g_{t\varphi}\Omega_{ob}-g_{\varphi\varphi}\Omega_{ob}^2\right]^{1/2}_{ob}}{\left[-g_{tt}-2g_{t\varphi}\Omega_{em}-g_{\varphi\varphi}\Omega_{em}^2\right]^{1/2}_{em}} \nonumber \\
& & \times\frac{\left(1+\Omega_{em}b\cos\eta\right)} {\left(1+ \Omega_{ob}b\cos\eta\right)}.
\label{redshift}
\end{eqnarray}

Finally, if we assume that the emitter and observer are in Keplerian rotation, then their angular velocities are given by
\begin{equation}
\Omega=\frac{M^{1/2}}{r^{3/2}+aM^{1/2}},
\label{omegakerr}
\end{equation}
with $r$ evaluated consistently with the observer and emitter coordinates. The above analysis reduces to the results of the Schwarzschild metric when we take the 
spin of the black hole $a=0$.

\begin{figure}[h]
\epsscale{1.}
\plotone{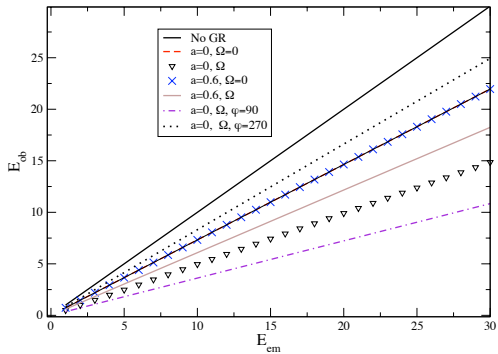}
\caption{Energy shifts for different black hole spin and rotation values compared to the non-relativistic case (No GR). The observer is located at $(r_{ob}=64.03 $ km, $\varphi_{ob}= 0,\theta=52.4^0)$. The emitter is at $(r_{em}=22.75$ km, $\theta=48.75^0$). Unless indicated otherwise, $\varphi_{em}=0$. $\Omega =0$ implies that the observer and emitter do not rotate around the black hole. The mass of the black hole is $2.5M_\odot$. The significant effect in the energy shift is due to the rotation of the emitter and observer around the black hole. \label{shiftplot}}
\end{figure}

In Figure \ref{shiftplot} we compare the energy shifts resulting from taking different values for the spin of the black hole with a mass of $2.5 M_\odot$ and 
allowing (or not) the emitter and the observer to orbit around it. In this example, we consider both observer and emitter located in the same plane $xz$ 
($\varphi=0$). We also examine the effect of rotation by comparing two cases.  We place emitters at the same distance $r_{em}$ but located at two different 
different angles $\varphi=90^0, 270^0$. In can be seen from this figure that there is not a significant difference between using a Kerr or a Schwarzschild metric 
when calculating the energy shifts if the spin of the black hole is moderate. A larger effect, however, comes from taking into account the possible rotation of the 
observer and emitter around the black hole.

%==========================================================================
\subsection{Fluxes}\label{fluxes}
We are interested in determining the neutrino fluxes observed at some point in the vicinity of a black hole. We start by determining the emitted fluxes from the 
neutrino surface. We assume a Fermi- Dirac distribution which is suitable for neutrinos. At the emission point this is
\begin{equation}
\phi(E_{em})=\frac{g_\nu c}{2\pi^2 (\hbar c)^3}\frac{E_{em}^2}{\exp(E_{em}/T_{em})+1},
\label{flux}
\end{equation}
with $g_\nu=1$, $T_{em}$ the temperature at the emission point, and with an assumed neutrino chemical potential $\mu_\nu=0$.
Using the known expression for the energy shifts, we can write this flux in terms of the energy measured by the observer
\begin{eqnarray}
\lefteqn{\phi_{em}(E_{em}=E_{ob}(1+z))= } \nonumber \\
& & \frac{g_\nu c}{2\pi^2 (\hbar c)^3}\frac{E_{ob}^2(1+z)^2}{\exp(E_{ob}(1+z)/T_{em})+1}.
\label{fluxemob}
\end{eqnarray}

The temperature measured by the observer relates to the temperature at the emission point by $T_{em}=(1+z)T_{ob}$. If we knew $T_{ob}$ we could replace this 
expression and then the shifts factors would cancel out in the exponential of equation \ref{fluxemob} to get
\begin{equation}
\phi_{ob}(E_{ob})=\frac{\phi_{em}(E_{em})}{(1+z)^2},
\label{fluxob}
\end{equation}
where 
\begin{equation}
\phi_{ob}(E_{ob})=\frac{g_\nu c}{2\pi^2 (\hbar c)^3}\frac{E_{ob}^2}{\exp(E_{ob}/T_{ob})+1},
\end{equation}
is the observed flux. 

In general, the observed effective flux coming from a finite source is
\begin{equation}
\phi^{\mathit{eff}}=\frac{1}{4\pi}\int d\Omega_{ob}\times\phi_{ob}(E_{ob}),
\end{equation}
where $d\Omega_{ob}$ is the solid angle that the source subtends as seen by the observer.  In this case one can use $\phi_{obs}$ from Eq. \ref{fluxob}.

Another way to see this is to start from the quantity $I/E^3$ which is an invariant \citep{gravitation}, where $I$ is the specific intensity, therefore we can 
calculate the observed flux in terms of the emitted flux as
\begin{eqnarray}
\phi^{\mathit{eff}}&\propto&\frac{1}{4\pi}\int d\Omega_{ob}\times\frac{I_{ob}(E_{ob})}{E_{ob}} = \nonumber \\
 & &\frac{1}{4\pi}\int d\Omega_{ob}\times\frac{I_{em}(E_{em})}{(1+z)^3E_{ob}},
\end{eqnarray}
where $I_{em}$ has the form, 
\begin{equation}
I_{em}\propto\frac{E_{em}^3}{\exp(E_{em}/T_{em})+1},
\end{equation}
because we have assumed a Fermi-Dirac distribution.

We would like to find reaction rates of neutrinos with matter around a black hole accretion disk. For this reason we want to calculate fluxes with energies in the 
observed system but using the values of emitted temperatures that come from the disk models. Therefore we make a change of variables back to the observed energy to 
finally get
\begin{equation}
\phi^{\mathit{eff}} \propto \frac{1}{4\pi}\int d\Omega_{ob}\times\frac{E_{ob}^2}{\exp(E_{ob}(1+z)/T_{em})+1},
\label{observedflux}
\end{equation}
for the observed effective flux. 

In order to determine $d\Omega_{ob}$ we assign a coordinate frame centered at the observer, located at a fixed point $r_{ob}$. Then $d\Omega_{ob}= \sin\xi d\xi 
d\alpha$, where $\xi$ and $\alpha$ the angles discussed in section \ref{neutrinotrajectories} and can be obtained by constructing null geodesics.  Note that these 
variables ($\alpha, \xi$) are different from the ones describing a coordinate system centered at the black hole.

\begin{figure}[h]
\epsscale{1.}
\plotone{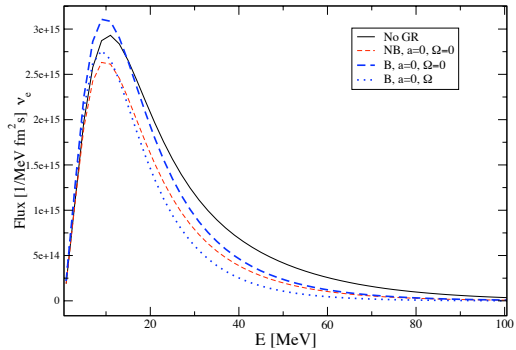}
\caption{Observed electron neutrino fluxes coming from a flat disk and observed at a point located in spherical coordinates $(r_{ob}=64.03 $ km, $\varphi_{ob}= 0,\theta=52.4^0)$. The mass of the black hole is $2.5M_\odot$. The solid line corresponds to the case where general relativistic effects are ignored and where the angular velocity of the disk $\Omega =0$ (No GR). The thin dashed line (NB) represents the flux when only the energy shifts were considered ($\Omega=0$). The bold dashed line (B) includes the energy shifts and the bending of neutrino trajectories. The dotted line includes all the effects plus the rotation of the observer and the emitter around the black hole. \label{fluxescompared}}
\end{figure}

In Figure \ref{fluxescompared} we compare the fluxes obtained when different general relativistic effects are considered. The plot shows electron neutrino fluxes 
from a thin flat disk (adapted from a 3D hydrodynamical model, as was used in \cite{Surman08}) and observed at a point with spherical coordinates $ (r_{ob}=64.03 $ 
km, $\varphi_{ob}= 0,\theta=52.4^0)$.  When only energy shifts are considered and we ignore rotation around the black hole, we find that the resulting neutrino 
fluxes are lower than the fluxes obtained for a non-relativistic case. However, taking into account the bending of neutrino trajectories makes the neutrino fluxes 
larger compared to the non-relativistic case for energies around 10 MeV, and larger compared to the case when the ray bending is ignored but the energy shifts are 
included.  On the other hand, the high energy talk of the distribution is reduced.  Adding the effect of rotation to the energy shifts decreases the flux at this spatial location in the model.  However, the effect of rotation can go in 
either direction, i.e. at some points it increases the flux and at some points it decreases the flux.

%==========================================================================
\section{Calculations of Nucleosynthesis from Black Hole Accretion Disks}
\label{nucleosynthesis}

We apply the general relativistic corrections to emitted neutrinos, as described in the previous section, from two 
specific models of black hole accretion disks.  The general strategy is to take calculated neutrino 
surfaces from the two different accretion disk models, and use these to find fluxes for points above the 
disk. Using these fluxes, we then find the nuclear products resulting from the interaction of the 
neutrino fluxes with material out-flowing the disk.

%========================================================================

%\subsection{Accretion Disk Models}
%\label{disk model}

We consider one dynamical and one steady state disk model. The first is based on a 3D hydrodynamical 
model of a black hole and a neutron star merger studied by Ruffert and Janka \citep{Setiawan04, Eberl99, 
Ruffert2001}. The second is based on a one dimensional steady state disk modeled by 
\citet{Beloborodovcross}.

The simulation by Ruffert and Janka corresponds to the merger of a $1.6 M_{\odot}$ neutron star and a 
$2.5 M_{\odot}$ black hole with spin parameter $a=0.6$. In this model general relativistic effects are 
included by using a modified Newtonian potential. The black hole is treated as a gravitational center 
surrounded by a vacuum sphere. The gravitational potential $\Phi_{BH}$ of the black hole is an extension 
of the Paczynski-Wiita potential \citep{BHpote} to a rotating black hole \citep{Artemova}.  As function 
of radius $r$, $\Phi_{BH}$ has the form
\begin{equation}
\frac{d\Phi_{BH}}{dr}=\frac{GM_{BH}}{r^{2\beta}(r-r_H)^\beta},
\end{equation}
where $\beta$ depends on the black hole spin parameter $a$, and $r_H$, $M_{BH}$ are the event horizon 
and mass of the black hole respectively. As a result of the coalescence, a disk is formed with inner 
boundary located at $\rho=14$ km and surface extending to $\rho=300$ km (in cylindrical coordinates).  
The model is dynamical and therefore the disk we consider is based on a snapshot of the simulation.
  
The model of \citet{Beloborodovcross} corresponds to a steady state disk. The mass of the black hole is 
3$M_\odot$, the accretion rate $\dot{M}=5M_\odot/$s, and the spin parameters used are $a=0$ and 
$a=0.95$. This model is fully relativistic. The disk is one dimensional, axially symmetric and is 
described by vertically averaged quantities. The disk extension goes as far as $\rho=600$ km. For the 
vertical structure of the disk we used a simple hydrostatic model that assumes an equilibrium with the 
gas radiation pressure and gravity.

%============================================================================
%\subsection{Calculations}\label{neutrinosurfaces}

\begin{figure}[h]
\begin{center}
\includegraphics[width=2.5in,angle=0,clip=true] {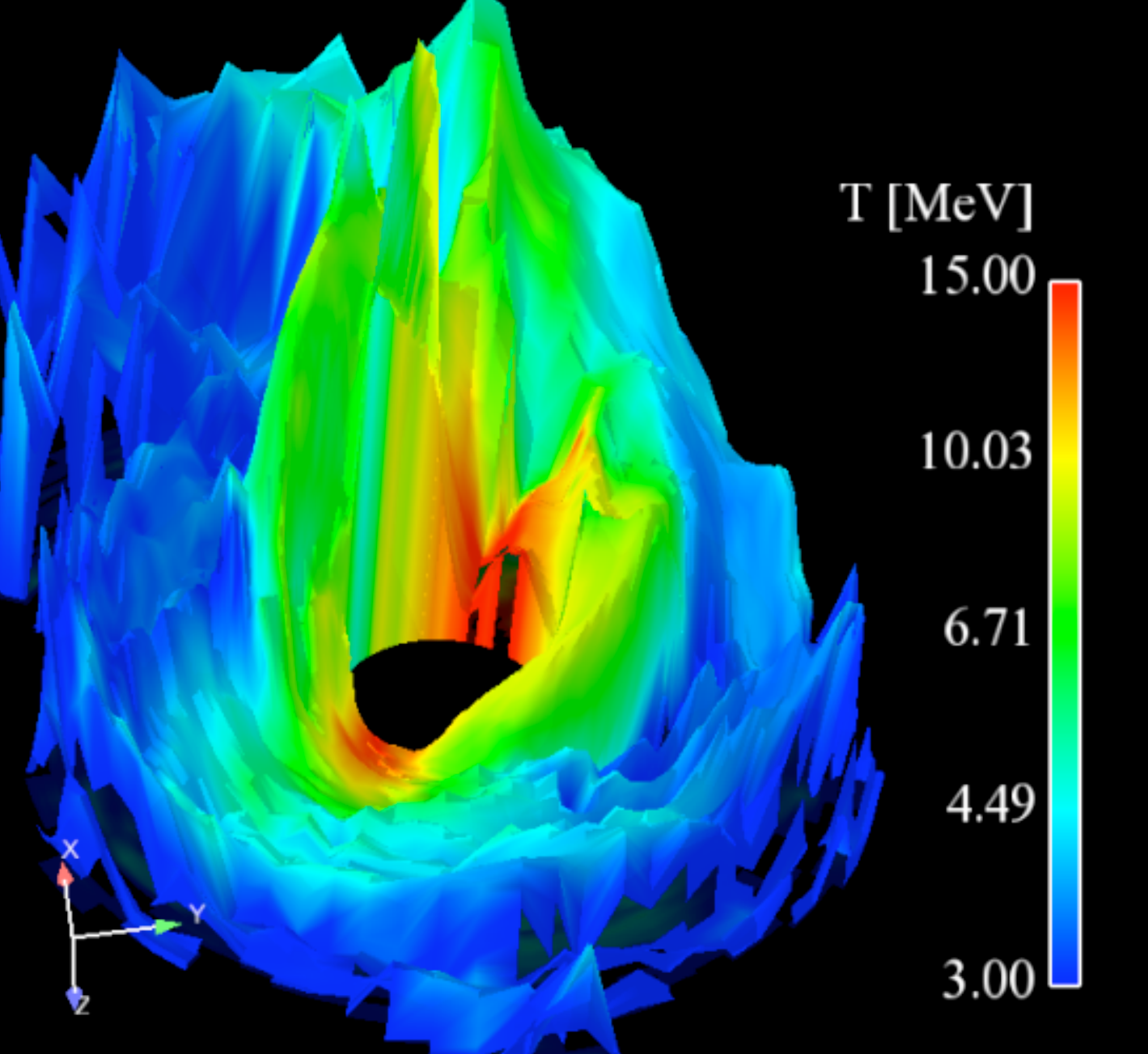}
\caption{Figure 3 in \citep{Caballerosurface}. Electron antineutrino surface seen at some inclination angle (see the $x$, $y$, $z$ axis on the lower left  corner). The height corresponds to the decoupling height $z_\nu$. The color scale corresponds to the emitted neutrino temperatures. The black area in the center represents the boundary with the black hole, $r=2r_s$. The neutrino surface shows a very uneven structure.}
\label{nuzT}
\end{center}
\end{figure}

In \citet{Caballerosurface} we calculated neutrino surfaces in the corresponding 3D cylindrical grids 
for both the steady state and the hydrodynamical models. Our results for neutrino surfaces for a fixed 
angle $\varphi$ can be seen in Figure 1 of \citet{Caballerosurface}. The neutrino surfaces found are not 
smooth. They present sharp variations in temperature, density and height. This fact can be seen in the 
3D image of the electron antineutrino surface showed in Figure 3 of \citet{Caballerosurface}. We 
replicate here this figure for illustration purposes (see Figure \ref{nuzT}). On the other hand the 
neutrino surfaces from the steady state Chen-Belobodorov model are smooth showing a torus shape.

When determining the trajectories of the neutrinos emitted from the disk, we consider two scenarios.  The 
first is the ``flat disk'' approximation.  In this approximation, while the neutrino temperatures are 
determined by the surface of last scattering, e.g. Figure \ref{nuzT}, the neutrino trajectories are 
started from the midplane ($z^\prime$=0) of the disk.  The second is the ``puffy disk'' where the neutrino 
trajectories begin at the surface of last scattering.

%\subsection{Outflow Model}
%\label{outflowmodel}

\begin{figure}[h]
\epsscale{1.0}
\plotone{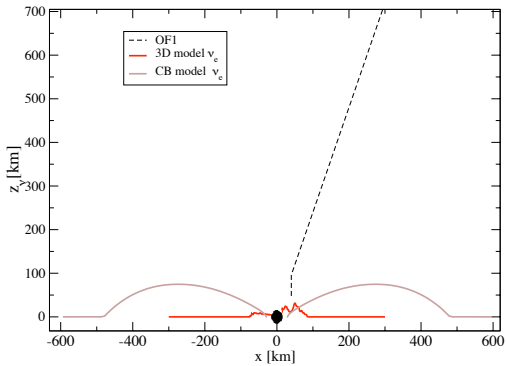}
\caption{A transversal view of the neutrino surfaces at $\varphi=0^\circ$ for the 3D hydrodynamical model(3D) and the steady state model (CB). The line OF1 shows the trajectory followed by the outflow. It starts in a straight line in the $z$ direction and then at the turnover point follow the radial direction. The black ellipse at ($x=y=0$) shows the black hole boundary.\label{fig:outflowsurface}}
\end{figure}

Since we are interested in the impact of the neutrino general relativistic corrections on 
nucleosynthesis, we must consider the path of ejected material from the disk.  Here we consider two 
types of outflow trajectories.  The first type is obtained with a similar prescription to that given in 
\citet{Surman05}, and used in \citet{Surman08}. In Figure \ref{fig:outflowsurface} we show this outflow 
trajectory and the electron neutrino surfaces for the steady state and hydrodynamical models. The image 
corresponds to a transversal cut at an angle $\varphi=0^\circ$. The trajectory starts at the point 
($\rho=40$, $z=48$) km and extends vertically up to a turnover point at ($\rho=40$, $z=96$) km, after 
which it follows the radial direction.  The outflow is taken to be adiabatic, with the velocity $v$ of 
the outflow as a function of distance from the black hole $r$ given by 
$v=v_{\infty}(1-r_{0}/r)^{\beta}$, where $r_{0}$ is the starting position on the disk, $v_{\infty}$ is 
the final coasting velocity of 0.1$c$, and $\beta$ determines the outflow acceleration, with lower 
$\beta$ corresponding to more rapidly accelerating outflows.  The second type of outflow trajectory 
considered is a parameterized spherical neutrino-driven wind trajectory, such as described in 
\citet{PanovJanka09}.  This trajectory is purely radial, starting at ($\rho=40$, $z=96$) km.  The 
outflow is again taken to be adiabatic, but with a homologous velocity-radius dependence ($v\propto r$) 
such that $r(t)=r_{0}e^{t/\tau}$, where $\tau$ is the dynamical timescale.  For both types of outflow 
trajectories, steady state conditions (i.e., $r^{2}\rho v=$ constant, where $\rho$ is the baryon 
density) are assumed.  More details of each outflow model can be found in \citet{Surman08} and 
\citet{PanovJanka09}.

We calculate the observed neutrino fluxes at every point of the outflow trajectories, coming from all 
the points of the neutrino surfaces showed in Figure \ref{nuzT}, and the corresponding torus-shaped 
neutrino surfaces of the steady state model.  In order to calculate the fluxes we follow the neutrino 
trajectories emitted from the coordinates ($\rho, \varphi, z_\nu$) (which can be translated to the 
spherical coordinates ($r_{em}, \varphi_{em}, \beta=\pi/2-\theta_{em}$) discussed in section 
\ref{neutrinotrajectories}), and arriving to the points of the outflow trajectory. Each point in the 
outflow trajectory has assigned coordinates $r_{ob},\varphi_{ob}, \theta_{ob}$. In this way we can 
calculate the angles $\xi$ and $\alpha$, formed in the sky of every point of the outflow, by neutrinos 
traveling from the neutrino surfaces, and therefore we determine the solid angle described by the disk 
at those points.  By means of Eq. \ref{observedflux} we calculate the corresponding fluxes. The energy 
shifts vary according to different conditions on the motion of the disk and the outflow, as well as the 
spin of the black hole.

Using these neutrino fluxes we calculate the element synthesis using a nuclear statistical equilibrium 
code, a charged particle reaction code, and, if necessary, an $r$-process network code, as described in 
\citet{Surman06}.  For the outflows from the asymmetrical 3D Ruffert and Janka disk, we take the 
additional step of following the outflow from starting points at four equally-spaced angles around the 
disk and then averaging the resulting abundance patterns.  We consider a number of cases to assess 
the impact on the abundance pattern of the general relativistic corrections to the neutrino flux.

%==============================================================================
\section{Results}
\label{results}

As material flows away from the disk, it begins as dissociated free neutrons and protons.  Then as the material cools, heavy 
nuclei are formed.  If the conditions are right then very heavy elements, such as the $r$-process elements or $p$-process elements, are 
formed. Of key importance in determining the type of nuclei formed are the relative numbers of neutrons and protons.  Neutrons are 
converted to protons through the weak interaction, in particular,
\begin{equation}
\nu_e + n \leftrightarrow p + e^-
\end{equation}
\begin{equation}
\bar{\nu}_e + p \leftrightarrow n + e^+
\end{equation}

We consider matter outflows that begin at the surface of a trapped source of neutrinos and end far from the source.  Close to the 
surface the neutron to proton ratio is determined by all four of these reactions.  However as the material flows away, there is a 
period where the most important reactions are the electron neutrino and electron antineutrino capture reactions.
\begin{figure}[h!]
\epsscale{1.0}
\plotone{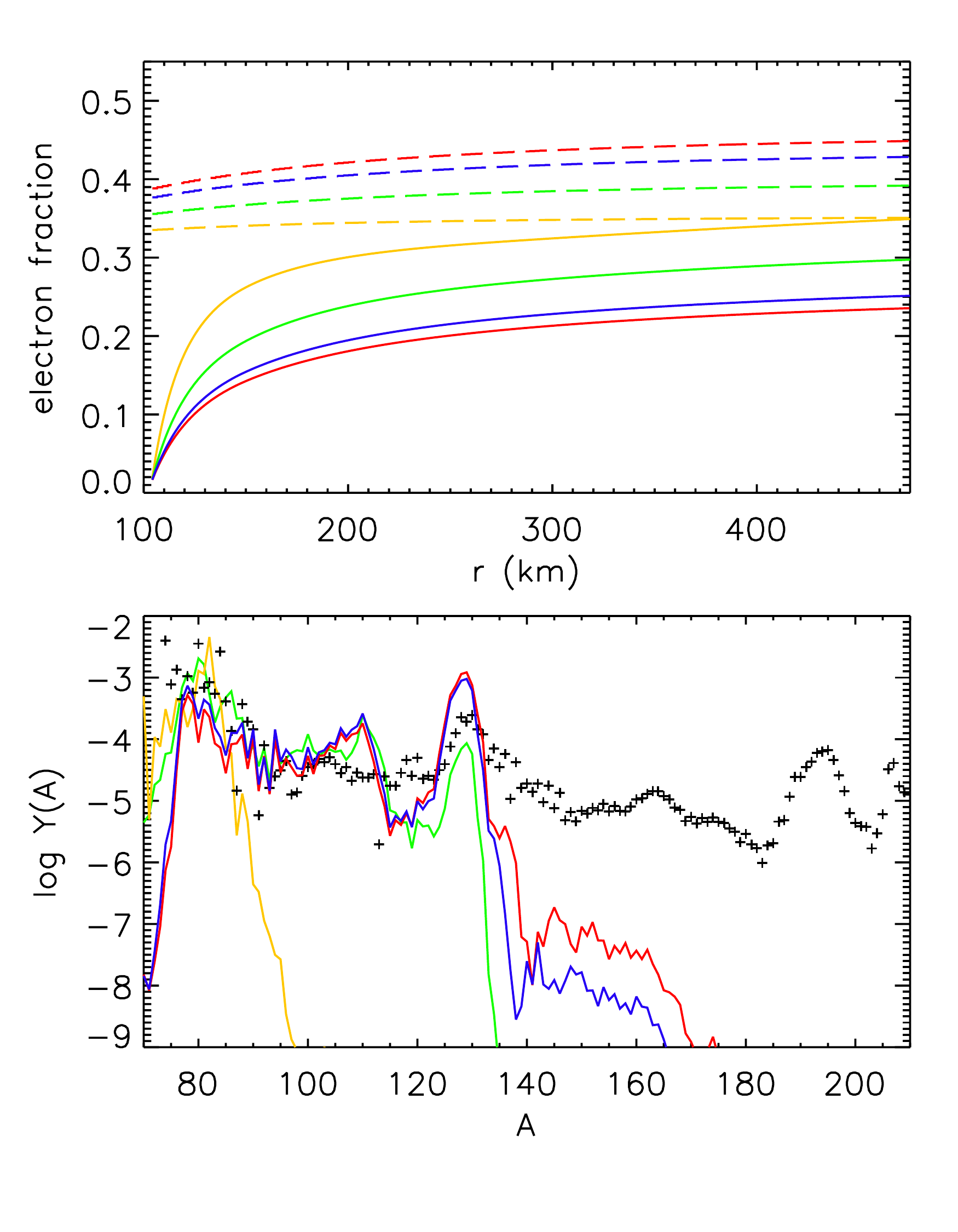}
\caption{The top panel shows electron fractions (solid lines) and neutrino equilibrium electron fractions (dashed lines) for the 
spherical wind trajectory with an entropy per baryon $s/k = 20$ and a timescale of $\tau = 5$ ms. In all cases the ``flat disk'' 
approximation to the black hole neutron star merger model was used so that all the neutrinos are launched from the z=0 plane.  The 
yellow lines treat the neutrinos as Newtonian. The green lines correspond to a situation where general relativistic neutrino 
redshift and trajectory bending are included, but rotation and black hole spin are not taken into account. The red lines add the 
effect of disk rotation.  The blue lines include a black hole spin of $a=0.6$ into the Kerr metric when calculating the 
gravitational redshift. The solid lines in the bottom panel show final nuclear abundances for each calculation, while the crosses 
show the scaled solar data.
\label{fig:flatdisk_fast}}
\end{figure}

In the top panel of Figure \ref{fig:flatdisk_fast}, one can see this effect in matter outflows from a black hole accretion disk.  
The solid lines show the electron fraction $Y_e = n_p / (n_p + n_n)$, where $n_p$ and $n_n$ are the number densities of the 
neutrons and protons in the material.  The dashed lines show the equilibrium electron fraction in the presence of only neutrinos, 
i.e. what the electron fraction would be if (1) the neutrino and antineutrino capture reactions were strong enough to establish an 
equilibrium between neutrons and protons and (2) electron and positron capture can be neglected. The different color lines 
correspond to different calculations of the neutrinos using the same disk model and matter outflow. (The differences in the 
calculations are explained below.)  It can be seen in the case of the yellow lines that the actual electron fraction approaches 
the equilibrium electron fraction at large distances.  In the case of the red, green and blue lines, the neutrino and antineutrino 
fluxes are lower and this equilibrium is not attained.

One can see also in this figure that the neutrino equilibrium electron fraction is not always the same.  This is due to the 
relative difference in the spectra of the neutrinos and the antineutrinos.  If the antineutrinos have higher energy than the 
neutrinos, then the material will be driven neutron rich, and vice versa.

In Figure \ref{fig:flatdisk_fast} we start with the black hole neutron star merger model and examine the simplest case: a flat disk.  When calculating the neutrino 
trajectories, we take the case of a black hole without spin, $a=0$. The choice of $a=0$ leads us to calculate the neutrino trajectories from the disk to the points 
of the outflow using the Schwarzschild metric as we discussed in section \ref{neutrinotrajectories}. We find the energy redshifts using Eq. \ref{sshift} of section 
\ref{energy red-shift}.  The impact of the neutrino gravitational redshift and neutrino trajectory bending can be seen in the figure.  For these calculations we 
have used a spherical neutrino-driven wind trajectory with a low entropy per baryon $s/k = 20$ and fast outflow ($\tau = 5$ ms). The yellow line shows the case 
with no neutrino general relativistic effects.  The neutrino and antineutrino fluxes are strong enough so that neutrino equilibrium of the electron fraction is 
obtained.  Furthermore, the antineutrino flux has higher energy than the neutrino flux and so the material remains neutron rich and the first peak of the $r$-process 
abundance pattern is obtained (yellow line bottom panel).

In the same figure, the green line shows a calculation where neutrino trajectory bending and neutrino redshift have been taken 
into account but without including the effects of the rotation of the disk. One can see that the solid and dashed lines never 
meet, and the equilibrium neutrino electron fraction is never obtained.  This is because both neutrino and antineutrino fluxes are 
less energetic when general relativistic effects are included.  In particular the high energy tails of the neutrino and 
antineutrino spectral distributions are reduced.

One can also see that the equilibrium electron fraction (green dashed line) is significantly higher than in the Newtonian case.  
This happens because the antineutrinos are on average emitted closer to the black hole than the neutrinos.  Thus the antineutrinos 
are more redshifted, and the balance of the weak rates moves toward proton rich material.  However, in this fast outflow scenario, 
the first effect dominates.  The equilibrium electron fraction is never obtained due to the overall decreased strength of the 
neutrino and antineutrino fluxes. Therefore the material retains much of its original neutron richness, and the $r$-process proceeds 
a little further - out to the second peak in this scenario (green line of the bottom panel).

The remaining two lines in Figure \ref{fig:flatdisk_fast} examine two additional effects.  The red lines include the effect of disk rotation in the 
 neutrino energy shifts, and the blue lines show 
%a black hole spin of $a=0.6$ in the redshift of the neutrinos using Eq. \ref{redshift} 
the influence of the black hole spin ($a=0.6$) on the redshift of the neutrinos (the Kerr metric was not used to compute null geodesics).  As can be seen from the figure, these effects create a more modest impact on the final abundance yields than the difference between 
treating the neutrinos as Newtonian or using the Schwarzschild metric to describe their evolution.

\begin{figure}[h]
%\epsscale{1.2}
\plotone{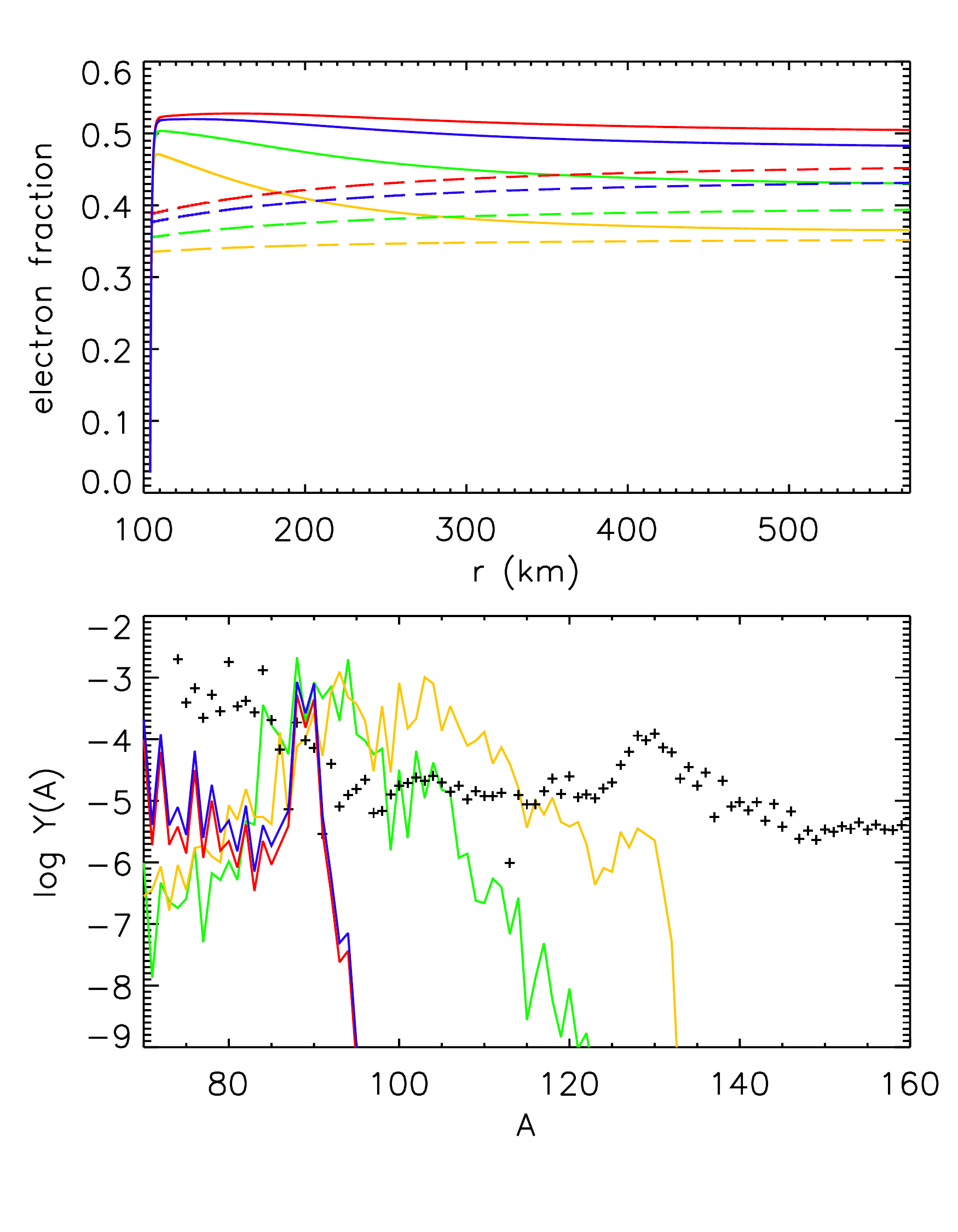}
\caption{The top panel shows electron fractions (solid lines) and neutrino equilibrium electron fractions (dashed lines) for the 
spherical wind trajectory with an entropy per baryon $s/k = 75$ and a timescale of $\tau = 50$ ms. 
In all cases the ``flat disk'' approximation was used so that all the neutrinos are launched from the z=0 plane.  The line colors 
represent the same cases as in Figure \ref{fig:flatdisk_fast}.
\label{fig:flatdisk_slow}}
\end{figure}

In Figure \ref{fig:flatdisk_fast}, much of the behavior of the abundance pattern shown in the bottom panel is due to the low entropy 
and fast outflow of the matter.  In Figure \ref{fig:flatdisk_slow} we show the same model with the one exception that we replace the 
matter outflow trajectory with one that has a higher entropy, $s/k = 75$, and slower outflow, $\tau = 50$ ms.  It can be seen that 
the initial electron fraction is much higher due to the higher entropy.  The final electron fraction in the Newtonian case is 
similar to the previous example since the neutrino and antineutrino fluxes are high enough for equilibrium to be reached.  In the 
cases where general relativistic effects are considered, again the actual electron fraction never obtains the equilibrium value. 
However, in this case, the electron fraction remains higher than the equilibrium value. While these high entropy matter outflows 
produce interesting nucleosynthesis, an $r$-process is not produced.

\begin{figure}[h]
%\epsscale{1.2}
\plotone{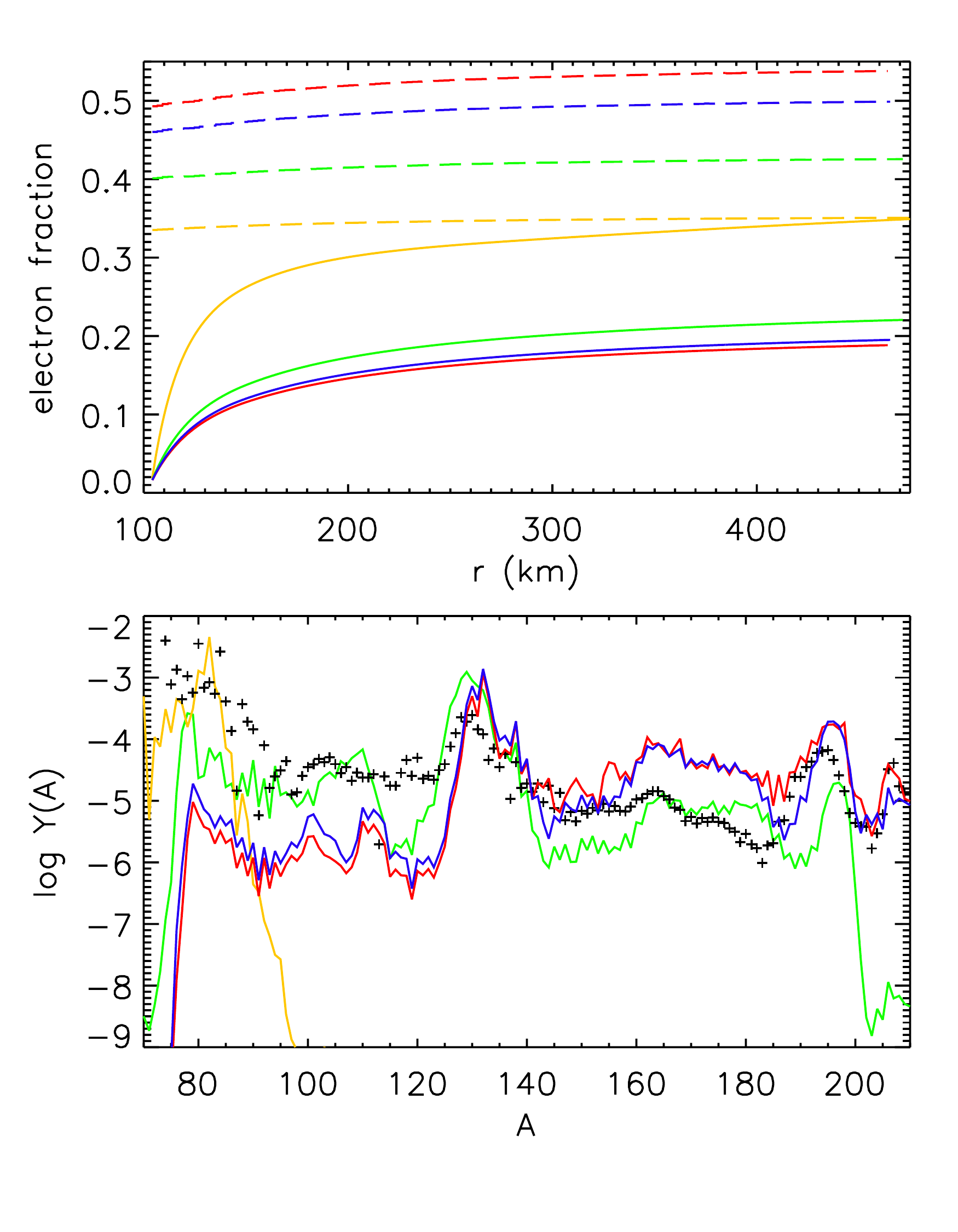}
\caption{ Same as Figure \ref{fig:flatdisk_fast} except that the ``flat disk'' approximation was not used.
\label{fig:puffydisk_fast}}
\end{figure}

When considering neutrino general relativistic effects, the emission point of the neutrinos is crucial.  Previous nucleosynthesis 
calculations from accretion disks have considered all neutrinos as if they were emitted from the $z^\prime=0$ plane of the disk, i.e. a 
flat disk approximation.  In Figure \ref{fig:puffydisk_fast}, we show the consequences of abandoning the ``flat disk'' 
approximation.  The same model and matter outflow trajectory was used as in Figure \ref{fig:flatdisk_fast}, but the ``flat disk'' 
approximation was not used and neutrino trajectories were started from the neutrino decoupling surface.

As can be seen from Figure \ref{fig:puffydisk_fast}, the behavior trend seen in the ``flat disk'' approximation has been accentuated 
in the ``puffy disk'' scenario.  The neutrinos are emitted further from the black hole and thus redshift effects are smaller, thus 
one might have expected the opposite behavior.  However, the hottest neutrinos are emitted on a surface that tilts toward the 
black hole, and so the difference in flux from geometrical considerations more than compensates for a softening of the redshift. 
We note also that the electron neutrino surface is higher than the electron antineutrino surface.  This enhances the flux of 
neutrinos relative to antineutrinos, and contributes to the proton rich equilibrium electron fractions seen in the top panel of 
Figure \ref{fig:puffydisk_fast}. Nevertheless, due to the overall reduction in strength of the neutrino and antineutrino fluxes, the 
material remains even more neutron-rich and a more robust $r$-process is obtained.

\begin{figure}[h]
%\epsscale{1.2}
\plotone{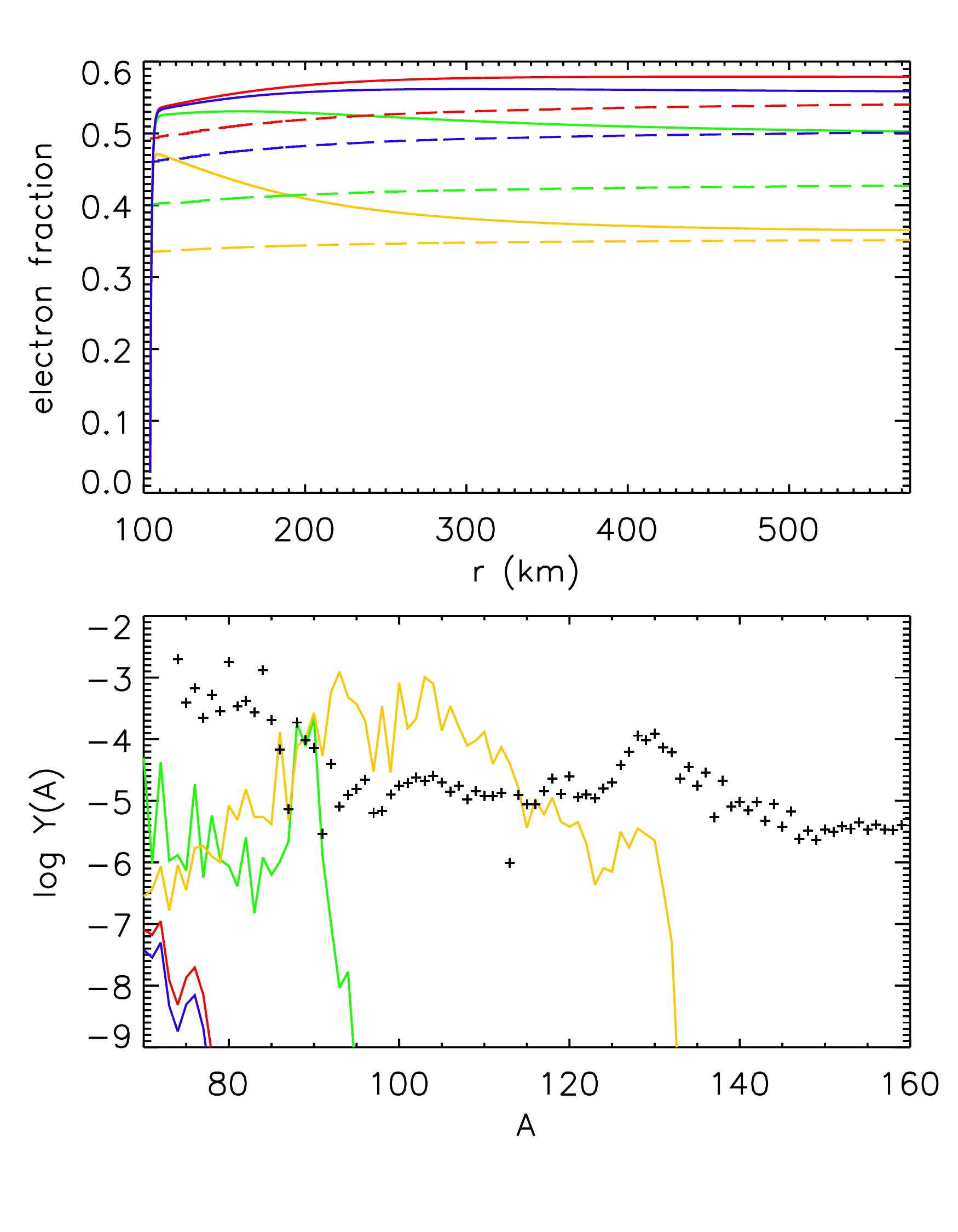}
\caption{ Same as Figure \ref{fig:flatdisk_slow} except that the ``flat disk'' approximation was not used.
\label{fig:puffydisk_slow}}
\end{figure}

\begin{figure}[h]
%\epsscale{1.2}
\plotone{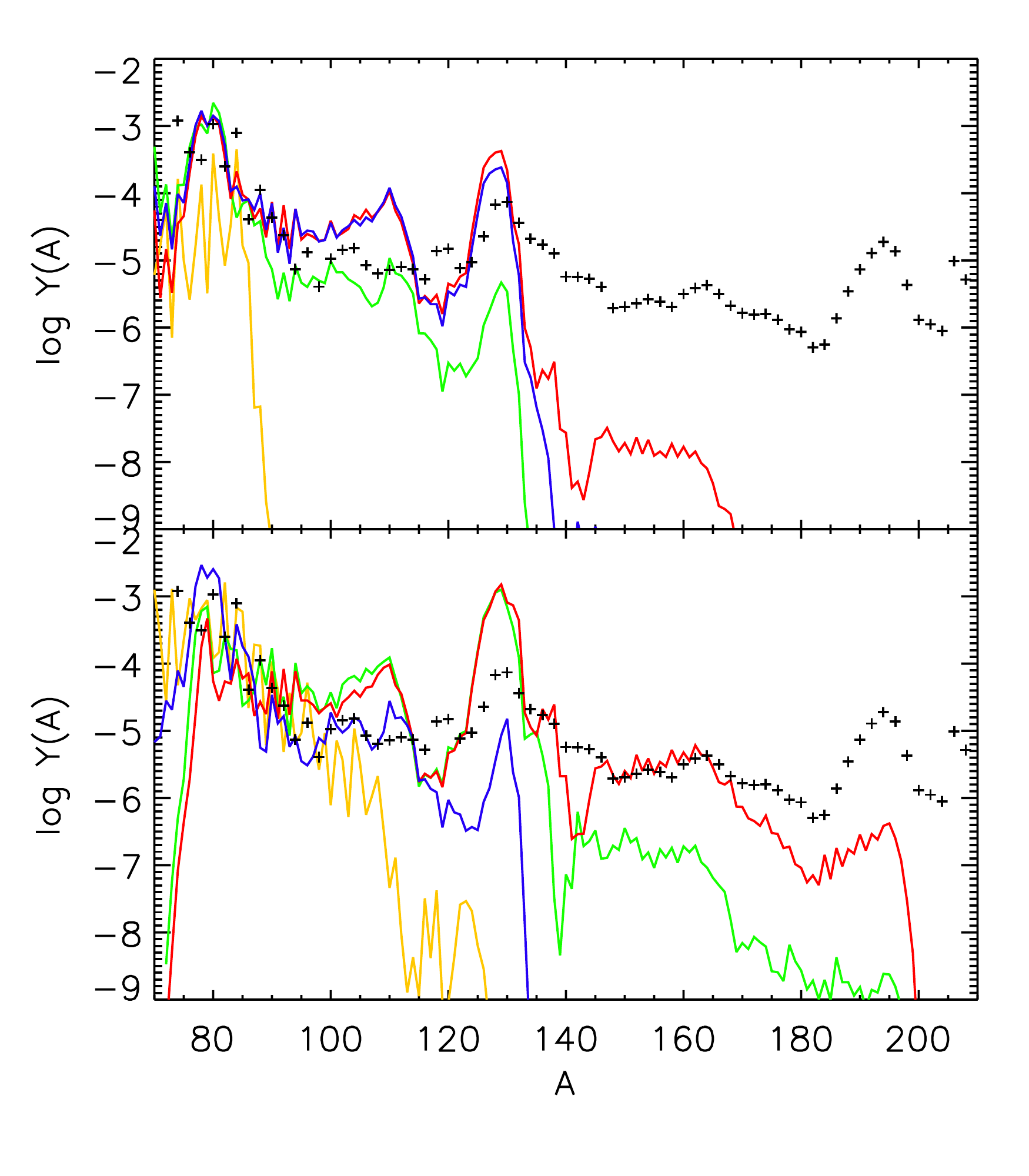}
\caption{Compares the results for the two different disk models using the outflow trajectory shown in Figure \ref{fig:outflowsurface} 
with an entropy per baryon $s/k = 10$ and acceleration parameter $\beta = 0.2$.  The colors indicate the same treatment of the 
neutrinos as in the preceding four figures, although for blue line in the bottom panel a black hole spin parameter of a=0.95 was used.  The top panel shows the results of a calculation which used the black hole neutron 
star merger disk and the bottom panel from a calculation which used the steady state disk.  The ``puffy disk'' was used in both 
panels.
\label{fig:comparison_fast}}
\end{figure}

If we instead compare the flat disk approximation, Figure \ref{fig:flatdisk_slow}, to the full ``puffy disk'' calculation, Figure 
\ref{fig:puffydisk_slow}, using high entropy and slower outflow conditions, we find that neutrino general relativistic effects can 
cause material that would be neutron rich in a Newtonian calculation to become proton rich.

To better understand whether these results are particular to a given disk and outflow model, we compare calculations done with the 
black hole neutron star merger model to those done with the steady state disk model.  We use the parameterized wind with the full 
outflow trajectory shown in Figure \ref{fig:outflowsurface}, a low entropy per baryon $s/k = 10$, and acceleration parameter 
$\beta=0.2$.  In Figure \ref{fig:comparison_fast} we show the abundance patterns obtained in this way.  The qualitative pattern seen 
in the results is similar to that seen in the low entropy cases of Figures \ref{fig:flatdisk_fast} and \ref{fig:puffydisk_fast}.

%=============================================================================

\section{Conclusions}\label{conclusions}

We have studied the influence of neutrino general relativistic effects on the spectra of neutrinos after they are emitted from black hole accretion disks.  We 
examined two models, a snapshot of a three dimensional dynamical calculation of a black hole neutron star merger and a steady state one dimensional disk.  We find 
that both redshift and trajectory bending are significant, and will influence not only the spectra of neutrinos but also nucleosynthesis products from material 
that is ejected from the disk near the source of neutrinos.

The overall impact of general relativistic effects on the neutrino spectra as compared with a Newtonian calculation is to reduce the energy flux of the neutrinos, 
and in particular to supress the high energy tails of the spectra.  Since the electron antineutrinos are emitted closer to the black hole then the electron 
neutrinos, the effect is stronger in this channel.  Thus, the electron antineutrino energy flux is reduced more than the electron neutrino flux.

Neutrinos are emitted from their surface of last scattering. The conditions at the point of last scattering initialize the neutrino spectral properties during free 
streaming phase.  From the point of view of determining the flux at every point above the disk (i.e. integrating over the emission surface), it can be 
computationally convenient to collapse the surface to the $z=0$ plane of the disk.  We compared two approaches, where neutrinos with the same spectral properties 
were emitted from either the plane of the disk (``flat disk'' approximation), or from the actual decoupling surface (``puffy disk'').  We found that the difference 
on the neutrino spectra above the disk was considerable, with the ``puffy disk'' scenario resulting in a lower energy flux that the ``flat disk'' scenario as 
calculated at the same point above the disk.  In the ``puffy disk'' scenario, there is less redshifting, since the neutrinos are emitted farther from the black 
hole.  However, the effect of the geometry of the disk is strong. The hottest neutrinos are emitted on a surface that is sloped toward the disk and the neutrinos 
bend toward the black hole, which reduces the fluxes of both neutrinos in directions away from the black hole, but has a larger impact on the antineutrinos.  Further, the neutrino emission surface is 
higher, so that the neutrinos are emitted closer to points of nucleosynthetic interest above the disk than the antineutrinos are.  This reduces the relative flux 
of antineutrinos relative to neutrinos at many points above the disk.

We considered the impact of these effects on the element synthesis in matter ejected from the disk in the vicinity of the neutrino decoupling surface.  The largest 
impact is in the reduction of the neutrino fluxes.  The neutrinos lose some of the influence that they had in setting the neutron to proton ratio.  Thus, the 
amount of heating that material receives as it leaves the surface of the accretion disk becomes an important factor in determining the final abundance pattern.  
In our models, with even modest heating and fast outflow, the neutron to proton ratio will be set almost entirely by electrons and positions, and only in the case 
of little to no heating will an $r$-process be produced.  In the case of slower outflow, the neutron to proton ratio will be set by the neutrino fluxes, but since 
the antineutrino flux is reduced more by general relativistic effects than the neutrino flux, the electron fraction of the material ranges from 0.4 to 0.6, which 
will produce interesting nucleosynthesis, but not an $r$-process.

Our study suggests that a range of nucleosynthesis products may be possible for accretion disks with trapped electron neutrinos and antineutrinos, and that the 
outcome is dependent on the neutrino spectra.  Neutrino spectra are sensitive not only to the general relativistic effects outlined here, but also to the 
decoupling surface of the neutrinos.  Therefore the nucleosynthetic outcome in ejecta from black hole accretion disks is as well. Thus, more detailed neutrino 
diffusion, preferably including the effects of general relativity, is warranted in future studies.

%\acknowledgments

\end{document}